\documentclass[11pt, letterpaper]{article} 

\usepackage[includeheadfoot,lmargin=1.2in,rmargin=1.2in,
            marginratio={1:1,2:3}, 
            height=688pt,]{geometry}

\usepackage{amsmath}
\usepackage{amsfonts}
\usepackage{amssymb}
\usepackage{bbm,bm}
\usepackage{mathtools}
\usepackage{slashed}
\usepackage{mathalfa}

\usepackage{graphicx,caption,subfigure}
\usepackage{tikz} 
\usepackage{tikz-cd}
\usetikzlibrary{decorations.pathreplacing,calc,tikzmark}
\usepackage{booktabs}
\usepackage{adjustbox}
\usepackage[utf8]{inputenc}
\usepackage[splitrule]{footmisc} 
 
\usepackage{placeins}
\usepackage{empheq}
\usepackage{paralist}
\usepackage{cite}
\usepackage[normalem]{ulem}
\usepackage{soul}

\usepackage[colorlinks=false,urlbordercolor=red
]{hyperref}
\usepackage{xcolor}
\usepackage{tensor}

\usepackage[none]{hyphenat} 
\newcommand{\nc}{\newcommand}
\nc{\lb}{\llbracket}
\nc{\rb}{\rrbracket}
\nc{\gl}{\llbracket}
\nc{\gr}{\rrbracket}
\nc{\del}{\partial}
\nc{\eq}[1]{\begin{equation}
                     \begin{split} #1 \end{split}
                     \end{equation}
}
\nc{\ov}{\overline}
\nc{\uli}{\underline}
\nc{\fa}{\hat}
\nc{\fb}{\MakeUppercase}
\nc{\fc}{\tilde}
\nc{\myhash}{\raisebox{\depth}{\#}}

\allowdisplaybreaks[2]
\numberwithin{equation}{section}
\DeclareUnicodeCharacter{2212}{-}

\newcounter{fneq} 
\counterwithin{fneq}{footnote}

\begin{document}


\vspace*{-1.5cm}
\begin{flushright}
  {\small
  MPP-2026-124
  }
\end{flushright}

\vspace{1.0cm} 
\begin{center}
  {\huge A Novel Matrix Model for the M5-brane?} 
\vspace{0.4cm}

\end{center}

\vspace{0.25cm}
\begin{center}
{
\Large Manuel Artime$^{1}$, Ralph Blumenhagen$^{1}$, and Thomas
Raml$^{1,2}$
}
\end{center}

\vspace{0.0cm}
\begin{center} 
  \emph{$^{1}$ 
Max-Planck-Institut f\"ur Physik (Werner-Heisenberg-Institut), \\ 
Boltzmannstra\ss e  8,  85748 Garching, Germany } 
\\[0.1cm] 
\vspace{0.25cm}
    \emph{$^{2}$ 
Arnold Sommerfeld Center for Theoretical Physics,\\
   Ludwig-Maximilians-Universit\"at, 80333 M\"unchen, Germany}
\vspace{0.3cm}
\end{center} 
\vspace{0.5cm}

\begin{abstract}
We provide a new formal extension of the BFSS matrix model by an
additional 5-bracket. Maximal supersymmetry leads us to promote the
BFSS 2-bracket structure constants to a dynamical field $H_{abc}$
governed by a Chern-Simons-like kinetic term. We first show that the
full generic pseudo-action is invariant under maximal supersymmetry
and that the associated supersymmetry algebra closes upon invoking a
number of BPS-like quadratic constraints. We then present a special
case based on the 5-Lie algebra $A_6$ that evades these additional constraints and allows for a Lagrangian description.
These results hinge on a conspiracy of properties of the $SO(9)$ gamma matrices and the 2- and 5-brackets. Compellingly, the resulting model seems to realize some features expected of a theory containing M5-branes and opens up the possibility of further including higher brackets for the M6- and M9-branes.
\end{abstract}

\thispagestyle{empty}
\clearpage

\setcounter{tocdepth}{2}

\tableofcontents

\section{Introduction}

A fundamental description of M-theory is still out of reach. The best we have so far is certainly the BFSS (Banks-Fischler-Shenker-Susskind) matrix model~\cite{Banks:1996vh} (see~\cite{Bilal:1997fy,Bigatti:1997jy,Taylor:2001vb} for reviews), which is believed to give an exact description of M-theory in the large-$N$ limit. Indeed, it was argued that the discrete light-cone quantization (DLCQ) of M-theory should be exactly described by the quantum mechanics of $N$ coincident D0-branes in Type IIA superstring theory~\cite{Seiberg:1997ad}.

By studying the interactions of bound states of such D0-branes~\cite{Kabat:1997im,Kabat:1997sa,VanRaamsdonk:1997in}, one could show that this theory also contains a longitudinal and a transverse M2-brane current, consistent with the earlier appearance of the same matrix model in the formulation of a discretized version of the membrane world-volume action~\cite{deWit:1988wri}.
However, the transverse M5-brane current did not appear at all and the longitudinal one was vanishing due to the Jacobi identity at finite $N$, so that one might wonder whether the BFSS model is really complete (see e.g.~\cite{Banks:1996nn,Berkooz:1996is,Castelino:1997rv,Taylor:1998tv} for some discussions on the role of M5-branes in BFSS and~\cite{Berenstein:2002jq, Maldacena:2002rb} for a deformation of the matrix model that realizes a certain configuration involving them). In accordance, recent Swampland considerations have revealed that in the decompactification limit to M-theory, both the M2- and M5-branes appear as light degrees of freedom~\cite{Blumenhagen:2023vhu}. Hence, they are on an equal footing, whereas the BFSS matrix model is biased towards M2-branes.

In this paper, we revisit this question and analyze whether an extension of the BFSS matrix model preserving maximal supersymmetry is possible. In view of the BLG (Bagger-Lambert-Gustavsson) theory~\cite{Bagger:2006sk,Gustavsson:2007vu,Bagger:2007jr,Bagger:2012jb}, which utilizes a three-algebra structure to describe the (leading order) theory on  multiple M2-branes (and an attempt for the world-volume theory of multiple M5-branes in~\cite{Lambert:2010wm}), we are led to consider an extension of the BFSS matrix model by a 5-bracket. Heuristically, one can think of the M2-brane of BLG as being described in matrix theory by a 2-bracket, so that to obtain the M5-brane one needs in total five transverse directions to the D0-branes, i.e. a 5-bracket.

A common objection to such a straightforward higher-bracket
implementation is that the M5-brane also carries a self-dual 3-form
field strength that needs to be taken into account. Nonetheless, we
take this as our starting point. As will be explained in the course of
this paper, the mathematical consistency of the model leads us to
introduce more structure, namely to promote the structure constants of
the BFSS 2-bracket to a dynamical field featuring a Chern-Simons-like
kinetic term.
Formulating a full-fledged supersymmetric theory out of these
ingredients is, however, a highly non-trivial task.
 In this article, we construct a supersymmetric model whose gauge symmetry is governed by a 5-bracket. 
Having multiple higher-brackets initially seems to require a set of so-called Filippov identities which turns out to be quite restrictive. Relaxing this requirement, we obtain a concrete solution presented in Section~\ref{sec:A6}. In a subsequent paper~\cite{Artime:toappear}, we clarify its connection to a quantum-mechanical model of the M5-brane and show how it encompasses a specific BFSS model.

Concretely, this paper is organized as follows: in
Section~\ref{sec:sec2}, we review the original BFSS matrix model and,
in particular, its symmetries. Then, in
Section~\ref{sec:BFSS_extension}, we introduce our new approach and
employ supersymmetry as a guide to uncover the structure of a new
quantum mechanical model involving both a 2- and a 5-bracket. 
We conclude with a discussion of some open questions and speculate on how the structure of the theory might extend to even higher brackets. Additional details of the computation are provided in the appendices.

\section{Review of the BFSS Matrix Model}\label{sec:sec2}

We briefly review the BFSS matrix model, focusing in particular on the way supersymmetry is realized. At first glance, the chosen formulation might seem unconventional, but it will serve as the basis for the generalization proposed in the remainder of this work.

\subsection{The supersymmetric BFSS matrix model}

The action of the BFSS matrix model is nothing but the dimensional reduction of 10D Super Yang-Mills theory with gauge group $U(N)$ to a single temporal direction. All fields are matrix-valued, and in the following we expand them into a  basis $T^a$, $a=0,1,\ldots, M$, with $M=N^2-1$,  of $U(N)$, where $T^0=\mathbb{I}/\sqrt{N}$ denotes the generator of the abelian $U(1)$ factor. Then, the degrees of freedom of this quantum mechanical model are
\begin{align}
    X^I(t) &= X^I_a(t) T^a\,,\qquad {\rm 9\ spatial\ bosonic\ fields\ with\ } I = 1, \dots, 9\,,\\
    A(t) &= A_a(t)  T^a\,,\qquad {\rm 1\ gauge\ field\ component}\,,\\
    \Theta(t)&=\Theta_a(t) T^a\,,\qquad \text{16-component real Majorana spinor}\,.
\end{align}
The generators satisfy the  standard normalization and commutation relations\footnote{Here and throughout the text, we will often write all gauge indices downstairs; in particular, this assumes the existence of a delta-like object to raise and lower indices.}
\begin{equation}
    \text{Tr}(T_a T_b) = \delta_{ab} \,,\qquad [T_a, T_b] = i f_{ab}{}^{c} T_c\,,
\end{equation}
where $f_{abc}$ are the totally antisymmetric structure constants of the gauge group. The non-dynamical gauge field $A(t)$ appears in the covariant derivative, which in components reads
\begin{equation}
    D_t X_a^I = \partial_t X_a^I + f_{abc} A_b X_c^I\,.
\end{equation}
Inspired by the BLG model~\cite{Bagger:2007jr} for stacks of M2-branes, we introduce a two-index gauge field
\begin{equation}\label{eq:defAab}
    \mathcal{A}_{ab}:= f_{abc} A_c\,,
\end{equation}
so that the covariant derivative becomes
\begin{equation}\label{eq:covderiv}
    D_t X_a^I = \partial_t X_a^I - \mathcal{A}_{ab} X_b^I\,.
\end{equation}
Note that $\mathcal{A}_{ab}$ is antisymmetric, which implies that $D_t$ satisfies all the relations expected of a covariant derivative. In these conventions, one can also define a gauge transformation $\Lambda_{ab}(t)= f_{abc} \lambda_c(t)$ acting as
\begin{equation}\label{eq:gaugeaction}
    \delta_{\Lambda} X^I_a = \Lambda_{ab} X^I_b \,,\qquad
    \delta_{\Lambda} \Theta_a = \Lambda_{ab} \Theta_b \,,\qquad
    \delta_{\Lambda} \mathcal{A}_{ab}=\partial_t \Lambda_{ab} + \Lambda_{ac} \mathcal{A}_{cb}+ {\Lambda}_{bc}  \mathcal{A}_{ac}\,.
\end{equation}
As employed in~\cite{Bagger:2007jr}, this allows one to treat the usual and higher-bracket gauge fields on an equal footing. The Jacobi identity for the structure constants implies that the usual matrix commutator
\begin{equation}\label{eq:lambdacomm}
    [\Lambda,\tilde\Lambda]_{ab}=\Lambda_{ac} \tilde\Lambda_{cb}-\tilde\Lambda_{ac} \Lambda_{cb}
\end{equation}
of two such gauge transformations results in another gauge transformation.

\noindent
Concerning the real 16-component Majorana fermions, the corresponding gamma matrices $\Gamma^I$  can be chosen to be real and symmetric, satisfying the Euclidean Clifford algebra $\{\Gamma^I, \Gamma^J\} = 2 \delta^{IJ} $ so that  one has $\ov{\Theta}_a=\Theta_a^T$. 

With these ingredients, the action of the BFSS matrix model reads
\begin{multline}\label{eq:actionbfss}
S_{\rm BFSS} =\! \frac{1}{2} \int  \mathrm{d}t \Bigl(  D_t X^I_a D_t X^I_a \!-\! \frac{1}{2} f_{abc} f_{ade} X^I_b X^J_c X^I_d X^J_e \!+\! i \Theta_a^T D_t \Theta_a \!+\! i f_{abc} \Theta_a^T \Gamma^I \Theta_bX^I_c \Bigr)\,.
\end{multline}
This action is invariant under time translations, the $SO(9)$ R-symmetry, shifts of $X^I_0$  and the gauge transformations defined above. For the latter one also needs to make use of the identity $\Lambda_{[\uli{a}d}f_{\uli{bc}]d}=0$ which when spelled out in terms of the definition of $\Lambda_{ab}$ is nothing but the Jacobi identity for $f_{abc}$.
In addition, this action is invariant under $16$ global, so-called dynamical, supersymmetry transformations acting on the fields of the BFSS matrix model as 
\begin{align}
\delta_\epsilon X^I_a &= -i \epsilon^T \Gamma^I \Theta_a\,,\\
\delta_\epsilon \Theta_a &= \left( D_t X^I_a \Gamma^I + \tfrac{1}{2}
  f_{abc} X^I_b X^J_c\, \Gamma^{IJ} \right) \,\epsilon\,,\\
\delta_\epsilon \mathcal{A}_{ab} &= -i \epsilon^T f_{abc}\Theta_c\,,
\end{align}
where $\epsilon$ is a constant Majorana spinor.
The invariance of the action under these transformations can be shown straightforwardly by using standard Clifford algebra relations. Importantly, one also needs the Fierz identity
\begin{equation}\label{eq:fierza}
\Gamma^I_{\alpha(\beta} \Gamma^I_{\gamma\delta)} =
\delta_{\alpha(\beta} \delta_{\gamma\delta)}\,,
\end{equation}
where $\alpha,\beta,\gamma,\delta\in\{1,\ldots,16\}$ denote spinor indices.

Note that the BFSS matrix model is invariant under additional $16$, so-called kinematic, supersymmetries, which only shift the fermion as
\begin{equation}\label{eq:kineticsusy}
  \delta_{\hat\epsilon} \Theta = \hat\epsilon\, \mathbb{I}\,,
\end{equation} 
leaving the other fields invariant. Notice that for the existence of these symmetries it is essential to have a $U(N)$ matrix model. On the component fields this symmetry simply  acts as $\delta_{\hat\epsilon} \Theta_0=\sqrt{N}\hat\epsilon$ leaving the other fermions invariant because of $f_{0bc}=0$. Hence, in total, the BFSS matrix model features 32 supersymmetries, which is the number expected for 11D M-theory. In the following we mainly focus on the non-trivial dynamical supersymmetries but come back to these kinematic ones in Section~\ref{sec_kinsusy}.

\subsection{Closure of the supersymmetry algebra}

If the action is invariant under supersymmetry, then it is also clear that $[\delta_1, \delta_2] S=0$, which means that  the commutator of two consecutive supersymmetry transformations must close into bosonic symmetries of the action. In the BFSS case these are time translations and gauge transformations, possibly extended by terms that vanish on-shell, i.e. upon invoking  the classical equations of motion.

In the BFSS matrix model, the algebra of dynamical supersymmetry transformations closes off-shell on the spatial coordinates and the gauge field
\begin{align}
    [\delta_{\epsilon_2}, \delta_{\epsilon_1}] X^I_a &= 2i \, (\epsilon_2^T \epsilon_1) \,D_t X^I_a + \delta_{\Lambda}  X^I_a\,,\\
    [\delta_{\epsilon_2}, \delta_{\epsilon_1}] \mathcal{A}_{ab} &=  D_t  \Lambda_{ab}\,,
\end{align}
with the gauge transformation
\begin{equation}
  \Lambda_{ab}=2i  (\epsilon_2^T \Gamma^I  \epsilon_1) f_{abc} X_c^I\,.
\end{equation}
For the real spinor field, the algebra closes only on-shell as 
\begin{equation}
\begin{aligned}
    [\delta_{\epsilon_2}, \delta_{\epsilon_1}] \Theta_a &= 2i \, (\epsilon_2^T \epsilon_1)\,D_t \Theta_a +\delta_{\Lambda} \Theta_a\\
    &-\tfrac{7}{8}i(\epsilon_2^T \epsilon_1)\, \rm{EoM}_\Theta -\tfrac{7}{8}i(\epsilon_2^T \Gamma^I \epsilon_1)\Gamma^I \, \rm{EoM}_\Theta +\tfrac{i}{192}(\epsilon_2^T \Gamma^{IJKL} \epsilon_1)\Gamma^{IJKL} \, \rm{EoM}_\Theta\,,
\end{aligned}
\end{equation}
where ${\rm{EoM}_\Theta} = D_t\Theta_a + f_{abc} \Gamma^I \Theta_b X^I_c$ is the equation of motion of $\Theta$.

One can also consider the closure involving the kinematic supersymmetry transformations. One finds that $[\delta_{\hat\epsilon_1}, \delta_{\hat\epsilon_2}] =0$ on all fields, and for the mixed case the only non-vanishing one is\footnote{Note that while the closure on $A_a$ gives a shift of the Abelian gauge field $A_0$, it vanishes on $\mathcal {A}_{ab}$.}
\begin{equation}
    [\delta_\epsilon, \delta_{\hat\epsilon}] X^I_a = i\sqrt{N}(\epsilon^T \Gamma^I \hat\epsilon)\, \delta_{a0}\,.
\end{equation}
Therefore, it closes up to a shift of $X^I_0$, i.e. the center of mass position of the D0-branes.

\subsection{Adding a 5-bracket}

We recall that the action of the BFSS matrix model is nothing but the DBI action for a stack of D0-branes
truncated to the quartic term. The question now is whether one can add more terms to the BFSS model to reconcile it with the results based on the M-theoretic Emergence Proposal~\cite{Blumenhagen:2023vhu}, which suggest that there should exist a theory that also includes transverse M5-branes in a more democratic manner.

To get an idea, we schematically express the bosonic part of the BFSS action in M-theory units
\begin{equation}
    S_{\rm BFSS}= \frac{1}{2R} \int \mathrm{d}t \,\mathrm{Tr}\bigg[ D_t X^I D_t X^I  + \frac{1}{2}\Big(M_*^3 R\, [X^I,X^J]\Big)^2 \bigg]\,,
\end{equation}
where $M_*$ is the eleven-dimensional Planck mass\footnote{To
  get~\eqref{eq:actionbfss} one needs to rescale $X'=M_* X$ and choose
  $M_*^2 R=1$.} and $R$ the size of the eleventh (light-cone) direction. Noting that the prefactor of the commutator term is essentially the tension of the M2-brane, the natural extension to include also the M5-brane is 
\begin{equation}
\begin{aligned}
    S= \frac{1}{2R}\int \mathrm{d}t \,\mathrm{Tr}\biggl[ D_t X^I D_t X^I &+\frac{1}{2}\Big(M_*^3 R\, [X^I,X^J]\Big)^2\\ &+ \frac{1}{5!}\Big(M_*^6 R\, [X^I,X^J,X^K,X^L,X^M]\Big)^2\biggr]\,,
\end{aligned}
\end{equation}
where, similar to the three-bracket in the BLG theory for stacks of M2-branes, we have introduced a 5-bracket. Recalling that in BFSS the transverse M2-brane charge is nothing but $Z^{IJ}=[X^I,X^J]$, it is natural to expect that the transverse M5-charge is then just given by the 5-bracket $Z^{IJKLM}=[X^I,X^J,X^K,X^L,X^M]$.

One might wonder where this new 5-bracket term comes from and why it has not been included before. Going to the Type IIA string frame, the action becomes
\begin{equation}
\begin{aligned}
    S= \frac{1}{2}\frac{M_s}{g_s} \int {\rm d}t \,{\rm Tr}\bigg[ D_t X^I D_t X^I & +\frac{1}{2}\Big(M_s^2\,[X^I,X^J]\Big)^2\\
    &+ \frac{1}{5!}\Big(\frac{M_s^5}{g_s}\, [X^I,X^J,X^K,X^L,X^M]\Big)^2 \bigg]\,,
\end{aligned}
\end{equation}
which makes it evident that the 5-bracket term cannot be seen in a weakly coupled string expansion and should be considered a genuine M-theory effect.  The remaining question is whether such a theory  can be constructed in a manner preserving its maximal supersymmetry, the $SO(9)$ R-symmetry, the time-translation and $X^I_0$-translation symmetries, while possibly generalizing the standard $U(N)$ gauge symmetry. 

\section{Extension of BFSS by a 5-bracket}\label{sec:BFSS_extension}

The purpose of this main section is to show that such a theory indeed exists. The main guiding principle will be the invariance of the action under the 16 dynamical supersymmetries.

\subsection{A first approach}\label{sec:first_approach}

First, we describe the new ingredients that we intend to add to the BFSS matrix model. The expectation is that the M5-brane can somehow be implemented by an antisymmetric 5-bracket
\begin{equation}
    [T_a,T_b,T_c,T_d,T_e]=F_{abcdef} T_f\,,
\end{equation}
which introduces a structure constant with six indices that, similarly
to a Lie algebra, we choose to be completely antisymmetric,
i.e. $F_{abcdef}= F_{[abcdef]}$. This will result in
  an algebraic structure that is not directly the $U(N)$ of BFSS. Recovering the usual BFSS model requires a more intricate treatment and will be addressed at a later stage. For example, one obtains
\begin{equation}
    [X^I,X^J,X^K,X^L,X^M]_f=F_{abcdef} X^I_a X^J_b X^K_c X^L_d X^M_e\,.
\end{equation}
Such a 5-bracket is usually subject to a quadratic relation, the fundamental identity, which is the generalization of the Jacobi identity for the standard 2-bracket. The fundamental identity, also called Filippov identity, of such a 5-bracket means that it acts on itself as a derivation\footnote{Concretely, this means
\vspace{-7pt}\begin{align*}
    &[ X^I,X^J,X^K,X^L,[X^M,X^N,X^P,X^Q,X^R]]= [[X^I,X^J,X^K,X^L,X^M],X^N,X^P,X^Q,X^R]\\[-2pt]
    &+[X^M,[X^I,X^J,X^K,X^L,X^N],X^P,X^Q,X^R]+\dots+[X^M,X^N,X^P,X^Q,[X^I,X^J,X^K,X^L,X^R]]\,.
\end{align*}}.
For the structure constants, this implies 
\begin{align}\label{eq:filippov55}
    F_{[\uli{abcde}}{}^k\, F_{k\uli{f}] lmnp} &=0\,,
\end{align}
where only the underlined indices are antisymmetrized over and
antisymmetrization, denoted by $[i_1\dots i_n]$, includes a factor of
$1/n!$. If, as in our case, a second bracket is present,
it is natural to impose
analogous conditions for an $m$-bracket to act as a derivation on the
$n$-bracket. For us this yields the four Filippov
identities
\begin{equation}
\begin{aligned}\label{eq:filipp}
    f_{[\underline{ab}}{}^k\, f_{k\uli{c}] l} &=0\,,\qquad (\rm{I})\\
    f_{[\uli{ab}}{}^k\, F_{k\uli{c}] mnp l} &=0\,, \qquad ({\rm II})\\
    F_{[\uli{abcde}}{}^k f_{k\uli{f}]g} &=0\,, \qquad ({\rm III})\\
    F_{[\uli{abcde}}{}^k\, F_{k\uli{f}] lmnp} &=0\,. \qquad ({\rm IV})   \,.
\end{aligned}
\end{equation}
As shown later, the status of the conditions we need to impose
is somewhat subtle. On the one hand, in Section~\ref{sec:revisit} we
will see that consistency with supersymmetry requires slightly
stronger conditions. On the other hand, the only known solution
to constraint (IV) is shown in Section~\ref{sec:A6} to yield a
supersymmetric theory without imposing constraints (I)–(III). However,
as a starting point, it is legitimate to develop the framework using these
identities.

As in the previous case, one can introduce a 5-bracket gauge field $A_{mnpq}$ and its gauge transformations. Again, it is enough to work with the combination
\begin{equation}\label{eq:25gaugefield}
    \mathcal{A}_{ab}= f_{abc} \,A_c + F_{abcdef}\, A_{cdef}\,,
\end{equation}
since the gauge fields enter the gauge transformation and covariant derivative only through $\mathcal{A}_{ab}$. This object plays the same role as the
  adjoint action of an ordinary Lie algebra gauge field, with the difference that its indices now label the vector space on which the 5-bracket acts, rather than the adjoint representation of $\mathfrak{u}(n)$. As a consequence, the parametrization~\eqref{eq:25gaugefield} carries some redundancy, which however plays no role in what follows, and one can work with $\mathcal{A}_{ab}$ directly.
  For now, however, we keep the split in terms of constituent fields as it highlights the similarity to existing constructions. We return to this point in Section~\ref{sec:A6}.
The corresponding gauge transformation reads
\begin{equation}\label{werder}
    \Lambda_{ab}= f_{abc}\, \lambda_c + F_{abcdef}\, \lambda_{cdef}\,.
\end{equation}
Then, the covariant derivative and the gauge transformations are still defined as before, i.e. in~\eqref{eq:covderiv} and~\eqref{eq:gaugeaction}. Moreover, the Filippov identities imply that the usual matrix commutator of two gauge transformations~\eqref{eq:lambdacomm} yields a gauge transformation of the form~\eqref{werder}. Given (I)--(IV) one can then compactly write
\begin{equation}\label{eq:filipp_Atilde}
   (a) \quad \mathcal{A}_{[\uli{a}}{}^{k}f_{\uli{bc}]k}=0\,, \qquad \qquad (b) \quad \mathcal{A}_{[\uli{a}}{}^{k}F_{\uli{bcdef}]k}=0\,.
\end{equation}
 
A first natural guess would be to extend the BFSS matrix model by the following higher-bracket interaction terms
\begin{equation}\label{eq:bfssextend}
\begin{aligned}
    S=S_{\rm BFSS}+\frac{1}{2} \int \mathrm{d}t \Big( &-\frac{1}{5!} F_{abcdef} F_{ab'c'd'e'f'} X^I_b X^J_c X^K_d X^L_e X^M_f\, X^I_{b'} X^J_{c'} X^K_{d'} X^L_{e'} X^M_{f'}\\
    &-\frac{i}{4!} F_{abcdef} \, \Theta_a^T \Gamma^{IJKL}\Theta_b  \, X^I_c X^J_d X^K_e X^L_f \Big)\,
\end{aligned}
\end{equation}
together with the extended supersymmetry transformations
\begin{align}
    \delta_\epsilon X^I_a &= -i \epsilon^T \Gamma^I \Theta_a\,,\\
    \delta_\epsilon \Theta_a &= \Big( D_t X^I_a \Gamma^I + \tfrac{1}{2} f_{abc} X^I_b X^J_c\, \Gamma^{IJ} + \tfrac{1}{5!} F_{abcdef} X^I_b X^J_c X^K_d X^L_e X^M_f \, \Gamma^{IJKLM} \Big) \,\epsilon\,,\\
    \delta_\epsilon \mathcal{A}_{ab} &= -i \epsilon^T \Big(f_{abc} -\tfrac{1}{3!} \Gamma^{IJK} F_{abcdef}   X^I_d X^J_e X^K_f \Big)\Theta_c\,,
\end{align}
where we have already fixed the numerical factors in hindsight. Also note that by introducing the gauge field $\mathcal{A}_{ab}$, we only have to specify its overall supersymmetry transformation. From this, one can then extract the supersymmetry transformations of the individual gauge fields $A_a$ and $A_{abcd}$ appearing in~\eqref{eq:25gaugefield}.

We have now arrived at a set of fields $\{X^I_a,\Theta_a,\mathcal{A}_{ab}\}$ whose dynamics is governed by an action and a set of transformations, so that we can proceed to check whether the action is invariant under supersymmetry transformations. Clearly, due to the high order of the added terms, one gets many new combinations of index contractions, involving also higher antisymmetric products of gamma matrices. In short, the system becomes quite complex.

To see whether there is a chance to cancel all terms, it is useful to isolate a term that only has a few contributions and is non-trivial to cancel. For the BFSS matrix model, this role is played by the fermionic cubic term, where one needs the Fierz identity~\eqref{eq:fierza} to cancel it. Therefore, we first look at the analogous highest fermionic order contributions. These arise from varying the gauge field $\mathcal{A}_{ab}$ in the kinetic term for the fermion $\Theta$ and from varying $X^I_a$ in the higher-order Yukawa-like term in~\eqref{eq:bfssextend}, so that
\begin{equation}
    \delta_{\epsilon} S \supset  \int \mathrm{d}t \Big(  \frac{1}{12}(\Theta_a^T \Theta_b)(\epsilon^T
    \Gamma^{JKL} \Theta_c) -\frac{1}{12}  (\Theta_b^T \Gamma^{IJKL}\Theta_c)(\epsilon^T
    \Gamma^{I} \Theta_a)\Big)  F_{abcdef} X^J_d X^K_e X^L_f \,.
\end{equation}
To cancel it, one needs a two-term Fierz identity for the involved gamma matrices. However, the relevant identity reads
\begin{equation}\label{eq:fierzb}
    \Gamma^I_{\alpha(\beta} \Gamma^{IJKL}_{\gamma\delta)} = \Gamma^{JKL}_{\alpha(\beta} \delta_{\gamma\delta)} -3 \Gamma^{[\uli{JK}}_{\alpha(\beta} \Gamma^{\uli{L}]}_{\gamma\delta)}\,,
\end{equation}
which evidently has one additional term. Therefore, as it stands, these cubic fermion terms do not cancel and we are facing an obstruction to supersymmetry.

\subsection{The supersymmetric extended BFSS model}

Closer inspection reveals that there is a chance to cancel this extra term by giving up the assumption that the structure constants $f_{abc}$ are actually constant and instead regard them as a dynamical field also transforming under supersymmetry. Equivalently, one could also add a dynamical contribution to the former structure constants, promoting them as
\begin{equation}
    f_{abc}\to f_{abc} +H_{abc}(t)\,.
\end{equation}
As we will see, it is tempting to consider  $H_{abc}$ as the BFSS realization of the self-dual three-form field strength for the  2-form living on the world-volume of an M5-brane\footnote{This should not be confused with the Kalb-Ramond two-form field $B$ and its three-form field strength $H=\mathrm{d}B$ appearing in the quantization of the closed string.}.
To ease the notation, in the following we simply write $H_{abc}$. To see that promoting the structure constants to a dynamical field results in the desired contribution, consider the Yukawa term in the BFSS action~\eqref{eq:actionbfss}
\begin{equation}
    S_Y= \frac{1}{2}\int {\rm d}t\,  iH_{abc}\, (\Theta_a^T \Gamma^L \Theta_b)  X^L_c \,.
\end{equation}      
If now $H_{abc}$ transforms under supersymmetry as
\begin{equation}\label{eq:Btransform}
    \delta_\epsilon H_{abc} = -\frac{i}{2} F_{abcdef} (\epsilon^T \Gamma^{JK} \Theta_d) X^J_e X^K_f\,,
\end{equation} 
then $\delta_\epsilon S_Y$ will cancel the third term induced by~\eqref{eq:fierzb}.

This observation turns out to be the key to consistently extend our first  naive proposal~\eqref{eq:bfssextend}. Having now a dynamical $H_{abc}(t)$ with a non-trivial supersymmetry variation, one has to go back and check what other terms are generated by varying it. Clearly, wherever $f_{abc}$ explicitly appeared in the action, one obtains new terms and one also has to be careful with partial integrations that can lead to new contributions involving $D_t H_{abc}$.

This happens in particular upon varying the kinetic terms for $X^I_a$ and $\Theta_a$ in the original BFSS model, which give rise to three contributions at order $H \Theta X D_t X$ which in BFSS canceled due to the relation $\Gamma^I \Gamma^J =\Gamma^{IJ} +\delta^{IJ}$. However, in bringing all terms to the same form, a partial integration has to be performed in one of the three contributions. For a dynamical field $H_{abc}$ this now gives an extra term
\begin{equation}\label{extravary}
    \delta_\epsilon S =\frac{i}{4} \int {\rm d}t \,  D_t H_{abc} \,(\epsilon^T \Gamma^{IJ} \Theta_a ) X^I_b X^J_c\,,
\end{equation}
which we need to cancel. We notice that the term that multiplies
\begin{equation}
  \label{werderB}
    D_t H_{abc}=\partial_t H_{abc} - 3\mathcal{A}_{[\uli{a} d} H_{d \uli{bc}]} 
\end{equation}
is very similar to the supersymmetry variation~\eqref{eq:Btransform} of $H_{abc}$ itself, although it does not coincide exactly due to the appearance of the 5-bracket structure constant. However, if we assume
that the latter ``squares to the identity''\footnote{Note
  that~\eqref{eq:squareid} is a prerequisite for defining an
  (imaginary) self-dual field $H_{abc}$ via $F_{abcdef} H_{def} =i H_{abc}$.}
\begin{equation}\label{eq:squareid}
    F_{abcmnp} \, F^{a'b'c'mnp}=\delta_{abc}^{a'b'c'}=\delta_{[\uli{a}}^{a'}\delta_{\uli{b}}^{b'}\delta_{\uli{c}]}^{c'} \,,
\end{equation}
then we can express~\eqref{eq:Btransform} as
\begin{equation}
    \frac{i}{2}(\epsilon^T \Gamma^{IJ} \Theta_{[a}) X^I_b X^J_{c]} =  F_{abcdef}  (\delta_\epsilon H_{def} )\,.
\end{equation}
Thus, we can cancel~\eqref{extravary} by adding a kinetic term
\begin{equation}\label{CSterm}
    S_{\rm CS}= \frac{1}{2}\int dt\,  H_{abc}\, F_{abcdef}\, D_t H_{def}
\end{equation}
to the action, which is reminiscent of a Chern-Simons-like (CS)
kinetic term. Like the latter, this term is not vanishing, since upon partial integration the reordering of the indices in $F_{abcdef}$ gives an extra minus sign.

Let us mention that the second term in the covariant derivative
\eqref{werderB} vanishes by~\eqref{eq:filipp_Atilde}, so that the covariant derivative reduces to the partial derivative. Analogously, the field $H_{abc}$ is invariant under gauge transformations.

So far, each step has required the introduction of further ingredients. In order not to lose track, let us summarize where we stand. In total, we have arrived at the action
\begin{equation}
\begin{aligned}\label{eq:bfssextendb}
    S= \frac{1}{2} \int \mathrm{d}t \bigg( &D_t X^I_a D_t X^I_a + i \Theta_a^T D_t \Theta_a +H_{abc}\, F_{abcdef}\, D_t H_{def}\\
    &- \frac{1}{2} H_{abc} H_{ade} X^I_b X^J_c X^I_d X^J_e +i  H_{abc} \Theta_a^T \Gamma^I \Theta_bX^I_c\\
    &-\frac{1}{5!} F_{abcdef} F_{ab'c'd'e'f'} X^I_b X^J_c X^K_d X^L_e X^M_f\, X^I_{b'} X^J_{c'} X^K_{d'} X^L_{e'} X^M_{f'}\\
    &-\frac{i}{4!}  F_{abcdef} \, \Theta_a^T \Gamma^{IJKL}\Theta_b  \, X^I_c X^J_d X^K_e X^L_f \bigg)\,
\end{aligned}
\end{equation}
and the supersymmetry transformations
\begin{align}
    \delta_\epsilon X^I_a &= -i \epsilon^T \Gamma^I \Theta_a\,,\label{eq:X_trafo}\\
    \delta_\epsilon \Theta_a &= \Big( D_t X^I_a \Gamma^I + \tfrac{1}{2} H_{abc} X^I_b X^J_c\, \Gamma^{IJ} + \tfrac{1}{5!} F_{abcdef} X^I_b X^J_c X^K_d X^L_e X^M_f \, \Gamma^{IJKLM} \Big) \,\epsilon\,,\label{eq:theta_trafo}\\
    \delta_\epsilon \mathcal{A}_{ab} &= -i \epsilon^T \Big(H_{abc} -\tfrac{1}{3!}  F_{abcdef}   X^I_d X^J_e X^K_f\Gamma^{IJK} \Big)\Theta_c\,,\label{eq:Atilde_trafo}\\
    \delta_\epsilon H_{abc} &= -\tfrac{1}{2}i F_{abcdef} (\epsilon^T \Gamma^{JK} \Theta_d) X^J_e X^K_f\label{eq:H_trafo}\,.
\end{align}
Additionally, from the action we obtain the following equations of motion
\begin{align}
    &\hspace{-19pt}\begin{aligned}\label{eoms1}
        0=&D_t^2X^I_a+H_{abc}H_{bde}X^J_cX^J_d\,X^I_e-\tfrac{i}{2} H_{abc}\Theta_b^T\Gamma^I\Theta_c \!+\! \tfrac{i}{2\cdot 3!}F_{abcdef} \Theta^T_b\Gamma^{IJKL}\Theta_c X^J_dX^K_eX^L_f\\
    &-\tfrac{1}{4!}F_{abcdef} F_{bghuvw} X^J_cX^K_dX^L_e X^M_f\,X^J_g X^K_h X^L_uX^M_v\,X^I_w\,,
    \end{aligned}\\    
    0=&D_t\Theta_a+H_{abc}\Gamma^I\Theta_bX^I_c-\tfrac{1}{4!}F_{abcdef}\Gamma^{IJKL}\Theta_bX^I_cX^J_dX^K_eX^L_f\,,\\
    0=&X^I_{[a} D_tX^I_{b]}-\tfrac{i}{2}\Theta_{[a}^T\Theta_{b]}-\tfrac{3}{2}H_{efg}F_{efg[\uli{a}cd}H_{\uli{b}]cd}\,,\label{eq:Gauss}\\
    0=&D_tH_{abc}-\tfrac{i}{2}F_{abcdef}\Theta^T_d\Gamma^I\Theta_e X^I_f+\tfrac{1}{2}F_{abcdeh}H_{fgh}X^I_dX^J_e\,X^I_fX^J_g\,,\label{eoms4}
\end{align}
where in order to derive the equation of motion for $H$ one needs to use~\eqref{eq:squareid}. The last terms of the equations of motion for $H$ and $\mathcal{A}$ vanish identically upon imposing the Filippov constraints
from~\eqref{eq:filipp}. We leave these terms written here explicitly, as they will play a role in Section~\ref{sec:A6}, but in the rest of the computations of this section we will set them to zero. The equation resulting from variation with respect to $\mathcal{A}$, \eqref{eq:Gauss}, is the generalization of the Gauss constraint in BFSS. We notice that the action continues to be invariant under time translations, the $SO(9)$ R-symmetry, and the usual gauge transformations. No new symmetry is apparent at this level.

Since a plethora of terms is involved, we have employed Mathematica\footnote{In particular, we have mostly used the package xAct~\cite{xAct}.} in order to simplify some expressions that appear in the supersymmetry transformations and subsequently checked whether the action is invariant upon invoking the four identities~\eqref{eq:filipp}. We can confirm that this is indeed the case, so that we have constructed a novel formal extension of the BFSS matrix model that preserves the $16$ dynamical supersymmetries.  We provide some more details of the invariance in Appendix~\ref{app_b}.

One might wonder how  this was possible by just introducing an additional  dynamical bosonic field $H_{abc}$ and no new fermionic superpartner. Recalling that the situation is very similar to ordinary Chern-Simons gauge theories in 3D, e.g. the BLG and ABJM (Aharony-Bergman-Jafferis-Maldacena)~\cite{Aharony:2008ug} theories, the resolution of this puzzle seems to be that $H_{abc}$ has no propagating degrees of freedom and hence is topological. Indeed, its CS-like kinetic term leads to an equation of motion of first order in time-derivatives and, in principle, does not lead to ``propagating wave'' solutions. In fact, as we demonstrate in Appendix~\ref{app:auxiliary_fermion} one can easily add a superpartner $\lambda_{abc}$ to the bosonic field $H_{abc}$, completing the multiplet. The equation of motion of this fermionic field is trivial, and integrating it out leads to the theory we have presented above.

\subsection{Closure of the supersymmetry algebra}\label{sec:closure}
 
Since we have constructed an action that is invariant under supersymmetry we expect the commutator of two supersymmetry transformations to close into time translations and gauge transformations, potentially upon invoking the equations of motion of some of the fields.

\noindent
For the closure on $X^I_a$ we get\footnote{We have observed that an analogous behavior does not hold for a similar extension of the BFSS model by a 3- or a 4-bracket. For an extension by an $n$-bracket, it only holds for $n(n+1)/2={\rm odd}$, cf. Appendix~\ref{app_b}.}
\begin{equation}
    [\delta_{\epsilon_2},\delta_{\epsilon_1}] X^I_a = 2i \, (\epsilon_2^T \epsilon_1) \,D_t X^I_a + \delta_{\Lambda} X^I_a\,,
\end{equation}
with the gauge transformation
\begin{equation}\label{eq:gaugeparam}
    \Lambda_{ab}=2i(\epsilon_2^T\Gamma^I\epsilon_1)H_{abc}X^I_c+2i(\epsilon_2^T\Gamma^{IJKL}\epsilon_1)\frac{1}{4!}F_{abcdef}X^I_cX^J_dX^K_eX^L_f\,.
\end{equation}
Hence, compared to BFSS, only the gauge transformation receives an additional term.

For the closure on the new field $H_{abc}$ we obtain
\begin{equation}\label{eq:closure_H}
    \begin{aligned}
    [\delta_{\epsilon_2},\delta_{\epsilon_1}] H_{abc} \!&=
    2i(\epsilon_2^T \epsilon_1) D_t H_{abc} +  2i(\epsilon_2^T \Gamma^I \epsilon_1) F_{abcdef} X^I_d \big( X^J_e D_tX^J_f -\tfrac{i}{2} \Theta^T_e \Theta_f \big)\\
    &-\! 2i(\epsilon_2^T \epsilon_1) \big(\!D_t H_{abc} \!-\!\tfrac{i}{2} F_{abcdef}\Theta_d^T \Gamma^J \Theta_e X^J_f\big),
\end{aligned}
\end{equation}
so that it closes up to the expected time derivative plus two terms that vanish upon invoking the equations of motion for the gauge field $\mathcal{A}_{ab}$ and $H_{abc}$.

The closure for the fermionic fields is more involved, but we managed to show that
\begin{equation}
\begin{aligned}
    [\delta_{\epsilon_2},\delta_{\epsilon_1}] {\Theta}_{a} &= 2i(\epsilon_2^T \epsilon_1) D_t \Theta_a + \delta_\Lambda \Theta_a\\
    &-\tfrac{7}{8}i(\epsilon_2^T \epsilon_1) \big( D_t\Theta_a + H_{abc} \Gamma^I \Theta_bX^I_c  -\tfrac{1}{4!}F_{abcdef} \Gamma^{IJKL} \Theta_b X^I_c X^J_d X^K_e X^L_f  \big)\\
    &-\tfrac{7}{8}i(\epsilon_2^T \Gamma^M\epsilon_1) \Gamma^M\!\big( D_t\Theta_a\! +\! H_{abc} \Gamma^I \Theta_bX^I_c \! -\!\tfrac{1}{4!}F_{abcdef} \Gamma^{IJKL} \Theta_b X^I_c X^J_d X^K_e X^L_f  \big)\\
    &+\tfrac{i}{192}(\epsilon_2^T\Gamma^{MNPQ} \epsilon_1)\\
    &\quad\times\Gamma^{MNPQ}\big( D_t\Theta_a + H_{abc} \Gamma^I \Theta_bX^I_c  -\tfrac{1}{4!}F_{abcdef} \Gamma^{IJKL} \Theta_b X^I_c X^J_d X^K_e X^L_f  \big)\,,
\end{aligned}
\end{equation}
i.e. it still closes up to a gauge transformation and the equation of motion for $\Theta_a$.

For the closure on the gauge field we obtain
\begin{equation}
  \label{closuregaugeA}
    \begin{aligned}
    [\delta_{\epsilon_2},\delta_{\epsilon_1}] \mathcal{A}_{ab} &= D_t \Lambda_{ab}\\
    &+\tfrac{i}{60} \, (\epsilon_2^T \Gamma^{IJKLM} \epsilon_1) H_{abc}  F_{cdefgh}  X^I_d X^J_e X^K_f X^L_g X^M_h\,,
\end{aligned}
\end{equation}
where the first term is the expected gauge transformation of $\mathcal{A}_{ab}$,
while the remaining term is somewhat unexpected. One interpretation is
that we might not yet have correctly implemented the gauge
symmetry. This is also supported by the special $A_6$ theory to be presented in Section~\ref{sec:A6}.
This theory will result in a supersymmetric theory without any constraints on 
 $H$, so that the latter  transforms covariantly under gauge
transformations. Indeed, the second term in the closure \eqref{closuregaugeA}
will turn out to be  absent in this theory.

Another possibility is that the extra term corresponds to a remnant gauge symmetry.
 For the moment, we notice that it can be written as
\begin{equation}\label{hannover96}
    [\delta_{\epsilon_2},\delta_{\epsilon_1}] {\cal A}_{ab}\supset H_{abc} \lambda_c\,.
\end{equation}
Recalling now the  initial definition of the two-index gauge field~\eqref{eq:25gaugefield}, one realizes that this corresponds to the transformation of the BFSS gauge field
\begin{equation}
    \delta_\lambda A_a = \lambda_a\,.
\end{equation}   
This is very reminiscent of the situation for the gauge field on a
D-brane, which is the combination ${\cal F}=B+ F$. Here $B$ is the
Kalb-Ramond 2-form gauge field from the bulk and $F=dA$ the field
strength of the 1-form gauge field on the brane. Then, the  1-form
gauge symmetry $\delta_\lambda B=d\lambda$  also acts like
$\delta_\lambda A=-\lambda$ on the gauge field. In this case, the
closure of the supersymmetry algebra acting on $A$ also involves on
the right hand side a gauge transformation $\delta_\lambda A$. We
propose that a very similar story can lie behind the appearance of the
extra term~\eqref{hannover96}. This appears to be a BFSS remnant of
the gauge transformation for the self-dual two-form on the M5
world-volume, which for consistency also acts on a one-form. 
See also~\cite{Pasti:1997gx} for a similar effect for the five-brane within 11D supergravity.

\subsection{Kinematic supersymmetries}\label{sec_kinsusy}

For the matrix theory to really provide a full description of 11D M-theory, it was essential that there also exist 16 kinematic supersymmetries acting as~\eqref{eq:kineticsusy}. The question, then, arises as to whether our model admits these extra supersymmetries as well.

As we have already discussed, for the matrix model it was crucial to have $U(1)\subset U(N)$ with $f_{0bc}=0$. Therefore, we need a distinguished index $a=0$ for which we also define $F_{0bcdef}$. Inspection reveals that the fermionic terms in the action~\eqref{eq:bfssextendb} can be made invariant under the kinematic supersymmetries by requiring $F_{0bcdef}=0$, which via~\eqref{eq:Btransform} implies that $\delta_{\epsilon} H_{0bc}=0$. One has now to analyze whether this spoils any of the steps we have performed in order to show the invariance of the action under dynamical supersymmetries.

A problem arises for the inversion of the relation~\eqref{eq:Btransform}, needed to obtain the CS kinetic term for $H_{abc}$. The resolution is that the component $H_{0bc}$ should not be considered as a dynamical field but rather be frozen to zero. To make this more transparent, we distinguish the index set $S=\{0,1,\dots,M\}$ from $\hat S=\{1,\dots,M\}$ and define
\begin{equation}
    F_{abcdef}=\begin{cases} F_{ABCDEF} & a,b,\ldots,f \in \hat S\\
    0 & {\rm else} \end{cases}\,.
\end{equation}
The dynamical fields are
\begin{equation}
    X^I_a\,, \Theta_a\,, H_{ABC}\,,
\end{equation}
with $H_{0BC}=0$ being non-dynamical, together with the gauge fields ${\cal A}_{ab}$ (which can be gauged away). For the action we then essentially take~\eqref{eq:bfssextendb} with indices $a,b,\dots\in S$, except for the CS term, which only involves the dynamical fields $H_{ABC}$, i.e.
\begin{equation}\label{eq:bfssextendcs}
    S_{\rm CS}= \frac{1}{2} \int {\rm d}t\, H_{ABC}\,  F_{ABCDEF}\,  D_tH_{DEF}  \,.
\end{equation}
Note that the four Filippov identities are still satisfied for the full index set $S$. Moreover, one can readily check that the closure involving also the kinematic supersymmetries gives the same result as for the BFSS matrix model, with the addition that the mixed commutator vanishes when acting on $H_{ABC}$. Therefore, the (pseudo-)action enjoys in total 32 supersymmetries, the number required for 11D M-theory.

\subsection{Revisiting the Filippov constraints}\label{sec:revisit}

Concerning the Filippov identities, one might ask what their status is, considering that the former structure constants $f_{abc}$ are now dynamical. First, we notice that the identities can be considered as BPS conditions on the polarized transverse M2- and M5-branes. The 5-bracket structure constants are non-dynamical and are to be considered as external parameters of the model, just like  the $f_{abc}$ in the original BFSS matrix model. In contrast, the other three Filippov identities involving $H_{abc}(t)$ are to be considered as additional constraints and should be implemented together with the equations of motion resulting from varying the (pseudo-)action with
respect to the dynamical field. However, one has to ensure that the set of constraints is compatible with the action of the symmetries. In the following, we first analyze these consistency conditions for the theory as defined by the set of four Filippov identities~\eqref{eq:filipp}. Subsequently, in Section~\ref{sec:A6} we show that under certain assumptions, which will single out the structure constant $\epsilon_{abcdef}$ of the 5-Lie algebra $A_6$, the set of constraints can be greatly relaxed, leading to a non-trivial space of solutions.

\subsubsection{Closure under supersymmetry}

Varying the usual Jacobi identity ${\rm (I)}$ and applying the Filippov identities~\eqref{eq:filipp} to simplify one of the resulting terms, one obtains the additional condition
\begin{equation}
  \label{eq:app2d}
    F_{[\uli{ab}def}{}^k\,  H_{k\uli{c}] l}\,\Gamma^{IJ} \Theta_d X^I_e
    X^J_f =0 \,.
\end{equation}
Similarly, the supersymmetry variation of relation ${\rm (II)}$ yields
\begin{equation}\label{eq:app2f}
    F_{[\uli{ab}def}{}^k\, F_{k\uli{c}]mnpl}\, \Gamma^{IJ} \Theta_d X^I_e
    X^J_f =0\,,
\end{equation}
while the variation ${\rm (III)}$ does not result in a new constraint.  Clearly, upon successive supersymmetry variations of these two constraints, one can generate a whole cascade of descendant constraints.

These  relations are not automatically satisfied by the equations of motion or the Filippov identities.  Thus, the original four Filippov identities~\eqref{eq:filipp} do not
manifestly close under supersymmetry. What one could do is to  strengthen the original identities and impose the set of Filippov-like constraints
\begin{equation}
\begin{aligned}\label{eq:strongconstraints}
    H_{[\underline{ab}}{}^k\, H_{k\uli{c}] l} &=0\,,\qquad (\rm{I})\\
    H_{[\uli{ab}}{}^k\, F_{k\uli{c}] mnp l} &=0\,, \qquad ({\rm II})\\
      F_{[\uli{ab}def}{}^k\,  H_{k\uli{c}] l}  &=0\,, \qquad ({\rm III'})\\
     F_{[\uli{ab}def}{}^k\, F_{k\uli{c}]mnpl} &=0\,, \qquad
     ({\rm IV'}) 
\end{aligned}
\end{equation}
all containing  just three terms. These are stronger in the sense that the original Filippov identities are implied by them. 
Note that, while the six-term Filippov identity $({\rm IV})$ was solved for the $SO(6)$ epsilon tensor $\epsilon_{abcdef}$ (up to some normalization factor), a non-trivial  solution to the stronger Filippov identity ${\rm (IV')}$ can be shown not to exist\footnote{This can be seen by picking $c = d$ and $(m,n,p,l)=(a,b,e,f)$ in the stronger Filippov identity ${\rm (IV')}$. We thank Jarod Hattab for pointing this out to us.}, rendering the theory essentially empty. Luckily, this is not the end of the story as it turns out that the initial four Filippov identities~\eqref{eq:filipp} can be further relaxed under certain conditions. This will be the topic of the next section.

\subsection[The \texorpdfstring{$A_6$}{A6}  extended BFSS model]{The \texorpdfstring{$\boldsymbol{A_6}$}{A6} extended BFSS model}\label{sec:A6}

For the fundamental identity underlying BLG \cite{Bagger:2006sk} it was proven, assuming the existence of a positive definite metric, that the only solution is the $3$-Lie algebra $A_4$ with the structure constants being the epsilon tensor of $SO(4)$ \cite{Figueroa-OFarrill:2002yjt,Papadopoulos:2008sk}. Here we can argue that, likewise, the only solution to the fundamental identity \eqref{eq:filippov55} is the epsilon tensor with indices in $SO(6)$. It is natural to ask whether, with this choice, the remaining Filippov constraints can be dropped, so that the failure of the constraint set to close under the symmetries of the action disappears and the physical content of the theory becomes easier to study. In the following we set $F_{abcdef}=\frac{1}{3!}\epsilon_{abcdef}$, hence the 5-Lie algebra $A_6$, and show that this is indeed the case.

The main difference with the previous treatment lies in \eqref{werderB}, where the Filippov identities (I) and (II) reduced the covariant derivative of $H_{abc}$ to a partial derivative. Now this is not the case anymore, and $H_{abc}$ is a gauge-covariant three-form transforming as
\begin{equation}
    \delta_\Lambda H_{abc}=3\Lambda_{[\underline{a} k}H_{k \underline{bc}]}\,.
\end{equation}
To show the invariance of the action, one needs to derive a set of
Schouten identities for the epsilon tensor, which hold because the
gauge indices run\footnote{The fields can be thought of transforming in the vector representation of $SO(6)$, in parallel with~\cite{Bagger:2007jr}.} only from $1$ to $6$. Concretely, we have 
\begin{equation}\label{eq:Schouten0}
    \delta_{n[\underline{m\vphantom{f}}}\epsilon_{\underline{abcdef}]}=0\,,
\end{equation}
so that contracting with $\Lambda_{nk}$ and setting $m=k$, one obtains
\begin{equation}\label{eq:Schouten1}
    \Lambda_{ak}\epsilon_{kbcdef}+\Lambda_{bk}\epsilon_{akcdef}+\Lambda_{ck}\epsilon_{abkdef}+\Lambda_{dk}\epsilon_{abckef}+ \Lambda_{ek}\epsilon_{abcdkf}+ \Lambda_{fk}\epsilon_{abcdek}=0\,.
\end{equation}
This suffices to prove that the action~\eqref{eq:bfssextendb} is gauge invariant\footnote{For the closure of the gauge-algebra of the extension presented before it was crucial that $H_{abc}$ and $F_{abcdef}$ satisfy the four Filippov identities~\eqref{eq:filipp} to show that after two successive gauge transformations the gauge parameter $\Lambda_{ab}$ is again of the form~\eqref{werder}. This is no longer required for the $A_6$ model due to the invertibility of $F \sim \epsilon_{abcdef}$, which spans all of the antisymmetric 2-forms $\Lambda^2 \mathbb{R}^6\cong \mathfrak{so}(6)$, and hence every antisymmetric gauge parameter $\Lambda_{ab}$ can be written as $\Lambda_{ab} \sim F_{abcdef}\lambda_{cdef}$.}.

Regarding invariance under supersymmetry transformations, since~\eqref{eq:Schouten0} contracted with $\mathcal{A}$ gives the analogue of~(b) in~\eqref{eq:filipp_Atilde}, most of the calculation remains unchanged. However, since (a) in~\eqref{eq:filipp_Atilde} does no longer hold, the variation of $D_tH_{abc}$ now reads
\begin{equation}
    \delta_\epsilon (D_t H_{abc}) = D_t \delta_\epsilon H_{abc} - 3i \epsilon^T \Theta_d H_{a[\uli{b}e}H_{\uli{cd}]e}- \frac{i}{2\cdot 3!}\epsilon^T \Gamma^{IJK}\Theta_d H_{[\uli{ab}h}\epsilon_{\uli{c}]defgh} X^I_e X^J_f X^K_g\,,
\end{equation}
where the latter two terms arise from the second term in~\eqref{eq:covDH}.
Thus, two new terms are generated by varying the CS term. The first of those can be shown to trivially vanish by noticing that an antisymmetrization over seven indices appears, while the indices only run from $1$ to $6$. In turn, the second term, after using \eqref{eq:squareid}, cancels exactly the variation of the $HHXXXX$ term in which the supersymmetry transformation acts on $X$,
\begin{equation}
    \delta_\epsilon S\supset -\frac{i}{2}\epsilon^T\Gamma^{IJK}\Theta_aH_{abe}H_{cd e} X_b^I X_c^J X_d^K\,,
\end{equation}
which previously vanished by the Filippov identity (I). Note that the discussion on kinematic supersymmetries of Section~\ref{sec_kinsusy} remains unaffected, since we can just append the index $0$ after considering $H_{0bc}=F_{0bcdef}=0$.

Two more terms were vanishing due to Filippov identities, namely
\begin{align}
    \delta_\epsilon S\supset &\frac{i}{6\cdot 3!}\epsilon^T\Gamma^{JKLM}\Theta_a\left(H_{bdh}\epsilon_{acefg h}-\tfrac{1}{4}H_{abh}\epsilon_{cdefg h}\right)X_c^I X_b^I X_d^J X_e^K X_f^L X_g^M\nonumber\\
    &+\frac{i}{48\cdot 3!}\epsilon^T\Gamma^{IJKLMN}\Theta_a\left(\tfrac{2}{5}H_{abh}\epsilon_{cdefg h} -H_{bch}\epsilon_{adefg h}\right)X_b^I X_c^J X_d^K X_e^L X_f^M X_g^N\,.
\end{align}
In the first line the two contributions are not identically zero, but can be shown to vanish after contracting \eqref{eq:Schouten1} with $H$, a choice of suitable indices, and imposing the antisymmetry in $[defg]$ and symmetry in $(bc)$. In the second line each term vanishes identically since antisymmetry in $[bcdefg]$ forces the contracted index $h$ to be an index already arising in $H$.

The equations of motion are \eqref{eoms1}--\eqref{eoms4}, with no
terms dropping out. Remarkably, a new term appears in the Gauss
constraint, which can be shown to be proportional to the former Jacobi
identity. Therefore, for static backgrounds with vanishing $\Theta$,
the Jacobi identity for $H$ is automatically implemented, while more
general backgrounds admit increasingly richer $H$ configurations. For
example, in \cite{Artime:toappear} we will present a non-collapsing $5$-sphere solution with
non-trivial time-dependent $H$ or rotating solutions with vanishing
$H$.
We thus interpret this Gauss constraint as enforcing the former BPS-like conditions dynamically. We leave a systematic study for future work.

\noindent
Finally, regarding the closure of the supersymmetry algebra, it is
straightforward to see that only $H$ and $\mathcal{A}$ get
affected. For $H$ one finds
\begin{equation}
    \begin{aligned}
    \!\!\!\!\![\delta_{\epsilon_2},\delta_{\epsilon_1}] H_{abc} \!&=
    2i(\epsilon_2^T \epsilon_1) D_t H_{abc}\\
    &+\!\frac{i}{3}(\epsilon_2^T \Gamma^I \epsilon_1) \epsilon_{abcdef} X^I_d \big( X^J_e D_tX^J_f -\tfrac{i}{2} \Theta^T_e \Theta_f -\tfrac{1}{4}H_{ghu}\epsilon_{ghuemn}H_{fmn}\big)\\
    &-\! 2i(\epsilon_2^T \epsilon_1) \big(\!D_t H_{abc} \!-\!\tfrac{i}{2\cdot 3!} \epsilon_{abcdef}\Theta_d^T \Gamma^J \Theta_e X^J_f\!\! +\!\!\tfrac{1}{2\cdot 3!} \epsilon_{abcdef} H_{dmn}  X^J_m  X^J_e   X^K_n  X^K_f \big)\\
    &+\!6i(\epsilon_2^T\Gamma^I\epsilon_1)H_{[\uli{a}de}H_{d\uli{bc}]}X^I_e+\tfrac{1}{4\cdot 3!}iH_{[\uli{bc}d}\epsilon_{\uli{a}]defgh}(\epsilon_2^T\Gamma^{IJKL}\epsilon_1)X^I_e X^J_f X^K_g X^L_h\,,
\end{aligned}
\end{equation}
where the last line is the expected gauge transformation with gauge parameter given by~\eqref{eq:gaugeparam}. Note that the terms quadratic in $H$ did not appear in the previous computation of the closure, which can be found in \eqref{eq:closureH}, but one can easily show that they give zero upon using \eqref{eq:squareid}.

In the closure of $\mathcal{A}$ there is a term that previously vanished
\begin{equation}
    [\delta_{\epsilon_2},\delta_{\epsilon_1}]\mathcal{A}_{ab}\supset \tfrac{1}{6}i(\epsilon_2^T\Gamma^{IJKLM}\epsilon_1)H_{cdh}F_{abefgh}X^I_cX^J_dX^K_eX^L_fX^M_g\,,
\end{equation}
which cancels against the gauge transformation given by $H_{abc}\lambda_c$ of \eqref{hannover96} after a series of Schouten identities. The rest of the terms remain unaffected.

Hence, by focusing on the case where
$F_{abcdef}\sim\epsilon_{abcdef}$, we obtain
a consistent model requiring no additional constraints on
$H_{abc}$.  As far as we know, this is the only solution to the fundamental
identity when requiring a positive definite inner product, anyway.

In the subsequent paper \cite{Artime:toappear}, we analyze this model in detail and show that it describes the quantum mechanics of a single M5-brane. Since our initial aim was to extend the BFSS matrix model, we also show there that the known BFSS Lagrangian is recovered in a particular scaling limit. The gauge algebra is, however, restricted to $\mathfrak{su}(2)\oplus\mathfrak{su}(2)$ such that the vector representation of $SO(6)$ decomposes into the adjoint of the two $SU(2)$ copies. This is a consequence of the lack of a parameter $N$. Recovering the full $U(N)$ symmetry of BFSS therefore remains an open challenge  and requires an extension of the ${\cal A}_6$ model to describe multiple M5-branes.

\vspace{1em}
\section{Conclusions}

In this paper we have presented a novel formal extension of the BFSS matrix model that involves a non-trivial 5-bracket as a new ingredient. The latter was motivated by our intention to make the M5-brane as explicit in the model as the M2-brane.

The guiding principle was the preservation of maximal supersymmetry, which turned out to be highly restrictive. In fact, it is very remarkable that such a consistent extension exists in the first place. This was only possible due to a constructive interplay of mathematical properties  of the $SO(9)$ gamma matrices and of the 2- and 5-brackets. The main new feature is that the structure constants of the 2-bracket  are treated as a dynamical field that is time-dependent and transforms non-trivially under supersymmetry.
Moreover, supersymmetry requires that its kinetic term is
Chern-Simons-like and that  the 6-index structure constant of the
5-bracket squares to one.
Hence, the extension we found shares remarkably many of the features
expected of a theory containing M5-branes. In particular, in the upcoming work \cite{Artime:toappear} we show that self-duality of $H$ is implemented naturally in the model, although the precise mechanism is not directly visible in the present discussion.

\paragraph{Open questions and future directions.}

Since we have promoted the former structure constants of the $U(N)$
Lie algebra to dynamical fields, the action~\eqref{eq:bfssextendb}
need not be a \textit{bona fide} matrix model in the standard
sense. Physically this means that the non-commutativity, or equivalently
the fuzziness, of the
coordinates becomes also dynamical, something one might expect
from a theory of quantum gravity.
The mathematical structure is rather reminiscent of the more general one of
$L_\infty$-algebras, but the quadratic constraints that we have are
actually much stronger than those of an $L_\infty$-algebra.
We leave a full clarification of the precise mathematical structure behind the model for future work.

One can recover the original BFSS matrix model by choosing a vanishing 5-bracket and considering $H_{abc}$ as non-dynamical. However, since the presence of a dynamical $H_{abc}$ is essential for the consistency of the new model, one cannot construct a supersymmetric theory with only a 5-bracket ---  at least not in the manner presented here. In physical terms, due to the existence of the 2-form on the M5-brane, one can always induce an M2-brane charge. 

On the other hand, as in the BLG theory, we have constructed a theory
with maximal supersymmetry at the expense of only having  a formal
theory in terms of  a 5-bracket and a dynamical $H_{abc}$  that are
subject to four (possibly stronger) Filippov identities plus the condition that
$F_{abcdef}$ squares to one. While the stronger constraints seem too
rigid to allow for a non-trivial solution, we have seen that the solution with $F_{abcdef}\sim\epsilon_{abcdef}$ leads to a supersymmetric theory without requiring the constraints on $H_{abc}$. Since with the current formulation this is the only solution to the Filippov identity (IV), studying the space of solutions of the unconstrained theory would serve as a first
approach  towards understanding what the more general theory describes.
In contrast, the lesson from the BLG and ABJM theories is that more solutions might be accessible by reducing the number of manifest supersymmetries or by weakening some of the conditions imposed so far on the 5-bracket. This might then allow to properly embed the construction in the full $U(N)$ symmetry of the BFSS model.

Finally, we cannot resist speculating on a possible even larger picture of extended BFSS matrix models.  As we  mentioned, the closure of two supersymmetry transformations acting on $X^I_a$ constrains the possible higher $n$-brackets to the ones where $n(n+1)/2$  is odd. Going through the list and taking into account that we have nine coordinates $X^I$, this leaves
\begin{equation}
         n=2:\  {\rm M2}\,,\qquad  n=5:\  {\rm M5}\,,\qquad   n=6:\  {\rm M6}\,,\qquad   n=9:\  {\rm M9}\,. 
\end{equation}
Quite remarkably, these are precisely the branes of M-theory, where M6
is the KK-monopole and M9 the Ho\v{r}ava-Witten domain wall. Hence, it is
tempting to speculate that in a next step one could also include an
additional 6-bracket in the construction, which again might lead to a
supersymmetric theory where now also the structure constants
$F_{abcdef}$ of the 5-bracket become dynamical. And then, as a final
step, one might even add a non-trivial 9-bracket to the system so that
also the structure constants of the former 6-bracket become
dynamical. Eventually the theory will only have the structure
constants of the 9-bracket   $F_{a_1\ldots a_{10}}$ as external  parameters with the remaining
structure constants all having become dynamical and subject to a long
list of BPS-like Filippov constraints. From this perspective, the
M5-brane theory discussed in this paper would  appear as  an
intermediate step.
It would be very satisfying 
if the step to M9 involved the hyperbolic Kac-Moody algebra
$E_{10}$ in some way \cite{Damour:2002cu}.

Perhaps such a complete particle-like theory could be considered as the  M-theory analogue of the Polyakov action for the string. In both cases,  the best available version of a  quantum gravity theory  would be  the background-dependent theory living on the lightest species  appearing in the respective infinite-distance limits of decompactifying (the eleventh direction) to M-theory and of a weakly coupled string theory.
Without having really constructed these additional maximally
supersymmetric extensions of the BFSS matrix model, these ideas are
certainly highly speculative, but maybe the insights from this paper
can serve as a guide to approach this problem.

Beyond the mathematical aspects, the most pressing question is what
this novel model actually describes, i.e. whether it correctly
implements M5-branes and their interactions in the discrete light-cone
quantization of M-theory. We leave these conceptual questions for
future studies \cite{Artime:toappear}.

\paragraph{Acknowledgments.}

We thank Antonia Paraskevopoulou for discussions and collaboration in the initial stages of the project.
We are also grateful to Niccol\`o Cribiori, Jarod Hattab and Carmine Montella for highlighting some aspects in an older version of the paper that needed further discussion, as well as to Matteo Zatti for insightful discussions that benefited the more complete current version.
We also thank the organizers of String Phenomenology 2026, where this work was first presented, and some of the participants for encouraging words and comments. The work of R.B. is supported  by the Deutsche Forschungsgemeinschaft (DFG, German Research Foundation) under Germany’s Excellence Strategy – EXC-2094 – 390783311.

\appendix
\addtocontents{toc}{\protect\setcounter{tocdepth}{1}}

\section{Gamma matrix relations and Fierz identities}\label{app_a}

In this appendix we collect some of the identities used throughout our computations.

We are working with gamma matrix conventions such that the transposition of an antisymmetrized product of $m$ gamma matrices satisfies:
\begin{equation}\label{gamma_transpose}
    (\Gamma^{I_1 \dots I_m})^T = (-1)^{\frac{m(m-1)}{2}} \Gamma^{I_1 \dots I_m}\,.
\end{equation}
Hence, $\Gamma^{IJ}$ and $\Gamma^{IJK}$  are antisymmetric while $\Gamma^I$, $ \Gamma^{I_1 \dots I_4}$ and $\Gamma^{I_1 \dots I_5}$ are symmetric matrices.

The product of two antisymmetrized gamma matrices is given by (see e.g.~\cite{Lauria:2020rhc})
\begin{equation}
    \begin{gathered}\label{eq:Gammaproduct}
    \Gamma_{I_1\dots I_n}\Gamma^{J_1\dots J_m}=\sum_{\substack{k=|n-m|\\k \text{ steps by }2}}^{n+m}\frac{n!m!}{s!t!u!}\,
    \delta^{[J_1}_{[I_n}\cdots\delta^{J_s}_{I_{t+1}}\,
    \Gamma_{I_1\cdots I_t]}{}^{J_{s+1}\cdots J_m]}\,,\\
    s=\frac{1}{2}(n+m-k)\,,\quad t=\frac{1}{2}(n-m+k)\,,\quad u=\frac{1}{2}(-n+m+k)\,,
\end{gathered}
\end{equation}
where we adopt the following conventions for the antisymmetrized gamma matrices,
\begin{equation}
    \Gamma^{IJ}=\frac{1}{2}(\Gamma^I\Gamma^J-\Gamma^J\Gamma^I)\,,
\end{equation}
and analogously for higher products.

Specializing to the Clifford algebra of $SO(9)$, for three general $16$-component Majorana spinors $\lambda\,,\psi\,,\chi$ the general Fierz rearrangement reads
\begin{equation}
    (\lambda^T\psi)\,\chi=-\frac{1}{16}\sum_{\Gamma}\pm_{\Gamma}(\lambda^T\Gamma\chi)\,\Gamma\psi\,,
  \end{equation}
where the sum runs over all independent antisymmetrized gamma matrices and $\pm_\Gamma$ is $+$ for the symmetric gamma matrices and $-$ for the antisymmetric ones. This identity allows us to reexpress the following contributions
\begin{align}
&\begin{aligned}
    &(\epsilon_2^T\Gamma\psi)\,\tilde\Gamma\epsilon_1-(\epsilon_1^T\Gamma\psi)\,\tilde\Gamma\epsilon_2=\\[0.1cm]&\quad-\tfrac{1}{8}\left((\epsilon_2^T\epsilon_1)\tilde\Gamma\Gamma\psi+(\epsilon_2^T\Gamma^I\epsilon_1)\,\tilde\Gamma\Gamma^I\Gamma\psi+\tfrac{1}{4!}(\epsilon_2^T\Gamma^{IJKL}\epsilon_1)\,\tilde\Gamma\Gamma^{IJKL}\Gamma\psi\right)\,,
\end{aligned}\\
&\begin{aligned}
    &(\epsilon_2^T\Gamma\psi)\,(\epsilon_1^T\tilde\Gamma\chi)-(\epsilon_1^T\Gamma\psi)\,(\epsilon_2^T\tilde\Gamma\chi)=\\&\quad\tfrac{1}{8}\left((\epsilon_2^T\epsilon_1)(\chi^T\tilde\Gamma^T\Gamma\psi)\!+\!(\epsilon_2^T\Gamma^I\epsilon_1)\,(\chi^T\tilde\Gamma^T\Gamma^I\Gamma\psi)\!+\!\tfrac{1}{4!}(\epsilon_2^T\Gamma^{IJKL}\epsilon_1)\,(\chi^T\tilde\Gamma^T\Gamma^{IJKL}\Gamma\psi)\right),
\end{aligned}
\end{align}
where $\Gamma,\tilde\Gamma$ can be any of the antisymmetrized gamma matrices. The resulting products can be further simplified using~\eqref{eq:Gammaproduct}.

\section{Computational details}\label{app_b}

In order to show the invariance of the action as well as the closure of the supersymmetry algebra, it is very convenient to first obtain the supersymmetry variation of the covariant derivatives
\begin{align}
    \delta_\epsilon (D_t X_a^I) &= D_t \delta_\epsilon X_a^I - X_b^I \delta_\epsilon\mathcal{A}_{ab}\,,\\
    \delta_\epsilon (D_t \Theta_a) &=D_t \delta_\epsilon \Theta_a - \Theta_b \delta_\epsilon \mathcal{A}_{ab}\,,\\ \delta_\epsilon(D_t H_{abc})&=D_t(\delta_\epsilon H_{abc})-3(\delta_\epsilon \mathcal{A}_{[\uli{a}d})H_{d\uli{bc}]}\,.\label{eq:covDH}
\end{align}
Upon (a) in~\eqref{eq:filipp_Atilde} the last line simply reduces to $\delta_\epsilon(D_t H_{abc})=D_t(\delta_\epsilon H_{abc})$. Furthermore, in the RHS of the equations there are terms arising that include
\begin{equation}
    D_t F_{abcdef}= \partial_t F_{abcdef} - 6 \mathcal{A}_{[\uli{a}g}F_{g\uli{bcdef}]}=0\,,
\end{equation}
which vanish identically by (b) in~\eqref{eq:filipp_Atilde}, just like their counterparts on the LHS that can be identified after properly reorganizing the relevant terms.

\subsection{Invariance of the action}

Performing a supersymmetry variation of the action with respect to the transformations~\eqref{eq:X_trafo}--\eqref{eq:H_trafo} and employing the gamma product relations~\eqref{eq:Gammaproduct}, we arrive at the rather lengthy expression
\begin{align}
    &\delta_\epsilon L=\nonumber\\
    &\frac{1}{2}(\epsilon^T\Theta_a)(\Theta_c^T\Theta_b)H_{abc} +\frac{1}{2}(\epsilon^T\Gamma^I\Theta_a)(\Theta_b^T\Gamma^I\Theta_c)H_{abc}\nonumber\\
    &-\frac{1}{4}(\epsilon^T\Gamma^{JK}\Theta_c)
    (\Theta_a^T\Gamma^I\Theta_b)F_{abcdef} X_d^I X_e^J X_f^K
    +\frac{1}{12}(\epsilon^T\Gamma^{IJK}\Theta_c)(\Theta_b^T\Theta_a)F_{abcdef}X_d^I X_e^J X_f^K\nonumber\\
    &-\frac{1}{12}(\epsilon^T\Gamma^I\Theta_a) (\Theta_b^T\Gamma^{IJKL}\Theta_c)F_{abcdef}X_d^J X_e^K X_f^L\nonumber\\
    &-\frac{i}{2}\epsilon^T\Gamma^{IJK}\Theta_aH_{abe}H_{cd e} X_b^I X_c^J X_d^K\nonumber\\
    &+\frac{i}{6}\epsilon^T\Gamma^{JKLM}\Theta_a\left(H_{bdh}F_{acefg h}-\tfrac{1}{4}H_{abh}F_{cdefg h}\right)X_c^I X_b^I X_d^J X_e^K X_f^L X_g^M\nonumber\\
    &+\frac{i}{48}\epsilon^T\Gamma^{IJKLMN}\Theta_a\left(\tfrac{2}{5}H_{abh}F_{cdefg h} -H_{bch}F_{adefg h}\right)X_b^I X_c^J X_d^K X_e^L X_f^M X_g^N\nonumber\\
    &+\frac{i}{12}\epsilon^T\Gamma^{LMN}\Theta_a F_{abdfh w}F_{ceguv w} X_c^I X_b^I X_e^J X_d^J X_g^K X_f^K X_h^L X_u^M X_v^N\nonumber\\
    &+\frac{i}{24}\epsilon^T\Gamma^{KLMNO}\Theta_a F_{abdfg w}F_{cehuv w} X_c^I X_b^I X_e^J X_d^J X_f^K X_g^L X_h^M X_u^N X_v^O\nonumber\\
    &-\frac{i}{144}\epsilon^T\Gamma^{JKLMNOP}\Theta_aF_{abdef w}F_{cghuv w} X_c^I X_b^I X_d^J X_e^K X_f^L X_g^M X_h^N X_u^O X_v^P\nonumber\\
    &-\frac{i}{2880} \epsilon^T\Gamma^{IJKLMNOPQ} \Theta_a F_{abcde w}F_{fghuv w} X_b^I X_c^J X_d^K X_e^L X_f^M X_g^N X_h^O X_u^P X_v^Q\nonumber\\
    &+\frac{i}{4} \epsilon^T\Gamma^{IJ}\Theta_a\left( D_t H_{abc} -F_{abcghu}F_{defghu}D_t H_{def}\right)X_b^I X_c^J\nonumber\\
    &+\frac{i}{2} \epsilon^T\Gamma^{IJ}\Theta_a \left(H_{def}F_{abcghu}F_{defghu} -H_{abc}\right)X_b^I D_t X_c^J\nonumber\\
    &+\frac{i}{4} \epsilon^T\Gamma^{IJ}D_t\Theta_c\left(H_{def}F_{abcghu}F_{defghu} -H_{abc}\right)X_a^I X_b^J\nonumber\\
    &+\frac{i}{48}\epsilon^T\Gamma^{IJKLM}\left(\Theta_a F_{abcdef} X_b^I X_c^J X_d^K X_e^L D_t X_f^M \!\!-\! \tfrac{1}{5} D_t\Theta_f F_{abcdef} X_a^I X_b^J X_c^K X_d^L X_e^M\right)\nonumber\\
    &-\frac{i}{2}\epsilon^T\Gamma^I D_t\Theta_a D_t X_a^I-\frac{i}{2}\epsilon^T\Gamma^I\Theta_a D_tD_t X_a^I\,.
\end{align}
First, it is easy to see that all terms involving covariant derivatives cancel exactly upon partial integration and use of the quadratic identity~\eqref{eq:squareid}. 
Second, the two terms in the first line vanish upon using the Fierz identity~\eqref{eq:fierza}, which is also needed for the standard BFSS matrix model. The three terms in lines 2 and 3 vanish precisely by use of the special Fierz identity~\eqref{eq:fierzb}.
Lastly, the remaining terms, which involve only a single fermionic bilinear, can all be shown to cancel or vanish as a consequence of the Filippov identities (I)--(IV).

\subsection{Closure of the supersymmetry algebra}
In the following we provide a few more details on the precise cancellations that lead to the closure of the supersymmetry transformations given in the main text.

For completeness, we compute the closure for $X^I$ when considering the introduction of a generic $n$-bracket. In particular, the supersymmetry transformation of $\Theta_a$ gets modified to
\begin{equation}
    \delta_\epsilon\Theta_a=\left(D_tX^J_a\Gamma^J+\frac{1}{n!}F_{aa_1\dots a_n}X^{J_1}_{a_1}\cdots X^{J_n}_{a_n}\Gamma^{J_1\dots J_n}\right)\epsilon\,.
\end{equation}
It is straightforward to show that
\begin{equation}
        [\delta_{\epsilon_2},\delta_{\epsilon_1}]X^I_a=2i(\epsilon_2^T\epsilon_1)D_tX^I_a+2i(\epsilon_2^T\Gamma^{J_2\dots J_n}\epsilon_1)\frac{1}{(n-1)!}F_{a a_1\dots a_{n}}X^{J_2}_{a_2}\cdots X^{J_n}_{a_n} X^I_{a_1}\,,
\end{equation}
after using the gamma matrix product
\begin{equation}
    \Gamma^I\Gamma^{J_1\dots J_n}=\Gamma^{IJ_1\dots J_n}+n\delta^{I[J_1}\Gamma^{J_2\dots J_n]}\,,
\end{equation}
and realizing that in order to obtain a suitable gauge transformation the term with $\Gamma^{IJ_1\dots J_n}$ needs to vanish. This occurs precisely for $n=2,5,6,9$ due to the symmetries of the gamma matrices.

For the remaining fields, we have performed the computations with a Mathematica notebook and now present the final cancellations. 
The closure for $\Theta$ reads
\begin{align}
    &[\delta_{\epsilon_2},\delta_{\epsilon_1}]\Theta_a=(\epsilon_2^{T}\epsilon_1)\left[\tfrac{9}{8}iD_t\Theta_{a} -\tfrac{7}{8}i H_{abc}\Gamma^{I}\Theta_{b}X^I_{c} 
    +\tfrac{7}{192}iF_{abcdef}(\Gamma^{IJKL}\Theta)_{b}X^{I}_{c}X^{J}_{d}X^{K}_{e}X^{L}_{f}\right] \nonumber\\[6pt]
    &+(\epsilon_2^{T}\Gamma^{I}\epsilon_1)\left[\tfrac{9}{8}iH_{abc}\Theta_{b}X^I_{c} 
    -\tfrac{7}{8}iH_{abc}\Gamma^{IJ}\Theta_{b}X^{J}_{c}+\tfrac{7}{192}iF_{abcdef}\Gamma^{IJKL}\Theta_{b}X^{I}_{c}X^{J}_{d}X^{K}_{e}X^{L}_{f}\right.\nonumber\\ &\left.\qquad\qquad\,\,\,\,\,+\tfrac{7}{48}iF_{abcdef}\Gamma^{JKL}\Theta_{b}X^I_{c}X^{J}_{d}X^{K}_{e}X^{L}_{f}-\tfrac{7}{8}i\Gamma^{I}D_t\Theta_{a}\right]\nonumber \\[6pt]
    &+(\epsilon_2^{T}\Gamma^{IJKL}\epsilon_1)\left[-\tfrac{1}{48}iH_{abc}\Gamma^{JKL}\Theta_{b}X^{I}_{c}\right. \!+\! \tfrac{1}{192}iH_{abc}\Gamma^{IJKLM}\Theta_{b}X^{M}_{c}\!+\!\tfrac{5}{64}iF_{abcdef}\Theta_{b}X^{I}_{c}X^{J}_{d}X^{K}_{e}X^{L}_{f}\nonumber\\
    &\qquad\qquad\qquad\quad-\tfrac{1}{48}iF_{abcdef}\Gamma^{ML}\Theta_{b}X^{M}_{c}X^{I}_{d}X^{J}_{e}X^{K}_{f}+\tfrac{1}{64}iF_{abcdef}\Gamma^{MNKL}\Theta_{b}X^{M}_{c}X^{N}_{d}X^{I}_{e}X^{J}_{f} \nonumber\\
    &\qquad\qquad\qquad\quad-\tfrac{1}{288}iF_{abcdef}\Gamma^{MNPJKL}\Theta_{b}X^{I}_{c}X^{M}_{d}X^{N}_{e}X^{P}_{f}\nonumber \\
    &\qquad\qquad\qquad\quad-\tfrac{1}{4608}i F_{abcdef}\Gamma^{MNPQIJKL}\Theta_{b}X^{M}_{c}X^{N}_{d}X^{P}_{e}X^{Q}_{f}\nonumber \\ 
    &\qquad\qquad\qquad\quad\left.+\tfrac{1}{192}i\Gamma^{IJKL}D_t\Theta_{a}\right]\,,
\end{align}  
The first line gives the expected time translation $2i(\epsilon_2^T\epsilon_1)D_t\Theta_a$ provided the following equation of motion is satisfied
\begin{equation}
    D_t\Theta_a + H_{abc} \Gamma^I \Theta_bX^I_c  -\frac{1}{4!}F_{abcdef} \Gamma^{IJKL} \Theta_b X^I_c X^J_d X^K_e X^L_f=0\,.
\end{equation}
The second group of terms can be simplified by using the gamma matrix products
\begin{equation}\label{eq:Gammainv}
    \Gamma^{IJ}=\Gamma^I\Gamma^J-\delta^{IJ}\,,\qquad \Gamma^{IJKLM}=\Gamma^I\Gamma^{JKLM}-4\delta^{I[J}\Gamma^{KLM]}\,,
\end{equation}
resulting in
\begin{equation}\label{eq:thetaclosure1}
    [\delta_{\epsilon_2},\delta_{\epsilon_1}]\Theta_a\supset (\epsilon_2^{T}\Gamma^{I}\epsilon_1)\left[2iH_{abc}\Theta_bX^I_c-\tfrac{7}{8}i\Gamma^I\cdot\mathrm{EoM}\,\Theta_a\right]\,.
\end{equation}
A similar computation is needed for the last group of terms, where one now needs the second identity from \eqref{eq:Gammainv} together with
\begin{align}
\begin{aligned}
    \Gamma_{IJKL}{}^{NOPQ}=&\Gamma_{IJKL}\Gamma^{NOPQ}-16\,\delta_{[L}^{[N}\,\Gamma_{IJK]}{}^{OPQ]}+72\,\delta_{[K}^{[N}\,\delta_{\vphantom{[}L}^{O\vphantom{[}}\,\Gamma_{IJ]}{}^{PQ]} \\
    &+96\,\delta_{[J}^{[N}\,\delta_{\vphantom{[}K}^{\vphantom{[}O}\,\delta_{\vphantom{[}L}^{\vphantom{[}P}\,\Gamma_{I]}{}^{Q]} -24\,\delta_{[I}^{[N}\,\delta_{\vphantom{[}J}^{\vphantom{[}O}\,\delta_{\vphantom{[}K}^{\vphantom{[}P}\,\delta_{L]}^{Q]}\,,
\end{aligned}
\end{align}
where we have written some indices downstairs to ease the notation. Remarkably, the remaining terms simplify to
\begin{equation}\label{eq:thetaclosure4}
    [\delta_{\epsilon_2},\delta_{\epsilon_1}]\Theta_a\supset (\epsilon_2^T\Gamma^{IJKL}\epsilon_1)\left[\tfrac{1}{12}iF_{abcdef}\Theta_bX^I_cX^J_dX^K_eX^L_f+\tfrac{1}{192}i\Gamma^{IJKL}\cdot\mathrm{EoM}\,\Theta_a\right]\,.
\end{equation}
The first terms in \eqref{eq:thetaclosure1} and \eqref{eq:thetaclosure4} combine into a gauge transformation of $\Theta_a$ with parameter
\begin{equation}\label{eq:gaugeparam_2}
    \Lambda_{ab}=2i(\epsilon_2^T\Gamma^I\epsilon_1)H_{abc}X^I_c+2i(\epsilon_2^T\Gamma^{IJKL}\epsilon_1)\frac{1}{4!}F_{abcdef}X^I_cX^J_dX^K_eX^L_f\,.
\end{equation}
Thus, the closure of $\Theta$ simplifies to
\begin{equation}
\begin{aligned}
    [\delta_{\epsilon_2},\delta_{\epsilon_1}]\Theta_a=&2i(\epsilon_2^T\epsilon_1)D_t\Theta_a+\delta_\Lambda\Theta_a\\&+\left[-\tfrac{7}{8}i(\epsilon_2^T\epsilon_1)-\tfrac{7}{8}i(\epsilon_2^T\Gamma^I\epsilon_1)\Gamma^I+\tfrac{1}{192}i(\epsilon_2^T\Gamma^{IJKL}\epsilon_1)\Gamma^{IJKL}\right]{\rm EoM}\,\Theta_a\,,
\end{aligned}
\end{equation}
so that, indeed, it closes on-shell up to time translations and gauge transformations. 

The closure for the $H$ field reads as follows
\begin{align}\label{eq:closureH}
    [\delta_{\epsilon_2},\delta_{\epsilon_1}]H_{abc} &=-(\epsilon_2^{T}\epsilon_1)F_{abcdef}(\Theta_d^{T}\Gamma^{I}\Theta_e)X^{I}_f-i(\epsilon_2^{T}\epsilon_1)H_{def}F_{abcghf}X^{I}_d X^{I}_g X^{J}_e X^{J}_h\nonumber\\[6pt]
    &+2i(\epsilon_2^{T}\Gamma^{I}\epsilon_1)F_{abcdef}X^{I}_d\!\left(X^{J}_eD_tX^{J}_f-\tfrac{i}{2}\Theta_e^{T}\Theta_f\right) \nonumber\\[6pt]
    &-\tfrac{1}{24}(\epsilon_2^{T}\Gamma^{IJKL}\epsilon_1)F_{abcdef}(\Theta_d^{T}\Gamma^{JKL}\Theta_e)X^{I}_f-\tfrac{3}{4}(\epsilon_2^{T}\Gamma^{I}\epsilon_1)F_{abcdef}\,(\Theta_d^{T}\Gamma^{IJ}\Theta_e)X^{J}_f \nonumber\\[6pt]
    &+\tfrac{1}{2}i(\epsilon_2^{T}\Gamma^{IJKL}\epsilon_1)H_{deh}F_{abcfgh}X^{I}_d X^{J}_e X^{K}_f X^{L}_g \nonumber\\
    &+\tfrac{1}{12}i(\epsilon_2^{T}\Gamma^{JKLMN}\epsilon_1)F_{abcdfw}F_{eghuv}{}_{w} X^{I}_e X^{I}_d X^{J}_f X^{K}_g X^{L}_h X^{M}_u X^{N}_v\,.
\end{align}
Adding and subtracting $2i(\epsilon_2^T\epsilon_1)D_tH_{abc}$ together with the terms appearing in the first line gives the expected time translation on $H_{abc}$, $2i(\epsilon_2^T\epsilon_1)D_tH_{abc}$, together with its equation of motion
\begin{equation}\label{eq:EoMH}
    D_tH_{abc}-\frac{i}{2} F_{abcdef}\Theta_d^T \Gamma^I \Theta_e X^I_f + \frac{1}{2}F_{abcdeh}H_{fgh} X^I_d X^J_e  X^I_f  X^J_g=0\,.
\end{equation}
The second line gives the equations of motion for $\mathcal{A}_{ab}$, which are identical to the BFSS ones
\begin{equation}
    X^{I}_{[a}D_tX^{I}_{b]}-\frac{i}{2}\Theta_{[a}^{T}\Theta_{b]}=0\,.
\end{equation}
The third line can be shown to vanish trivially as the gamma matrices with two and three indices are antisymmetric, while the last two lines are also zero by using the Filippov identities (I)--(IV) provided in Section~\ref{sec:first_approach}.
Finally, the closure for the gauge field $\mathcal{A}_{ab}$ is given by
\begin{align}
[\delta_{\epsilon_2},\delta_{\epsilon_1}]\mathcal{A}_{ab} &=2i(\epsilon_2^{T}\Gamma^{I}\epsilon_1)H_{abc}D_tX^{I}_c+\tfrac{1}{3}i(\epsilon_2^{T}\Gamma^{IJKL}\epsilon_1)F_{abcdef}X^{I}_c X^{J}_d X^{K}_eD_tX^{L}_f\nonumber\\[6pt]
    &-2i(\epsilon_2^{T}\Gamma^{I}\epsilon_1)X^I_c \left[-\tfrac{1}{2}i F_{abcdef}\Theta_d^T \Gamma^J \Theta_e X^J_f + \tfrac{1}{2}F_{abcdef}H_{dmn}  X^J_m  X^J_e   X^K_n  X^K_f \right]\nonumber\\[6pt]
    &-\tfrac{1}{60}i(\epsilon_2^{T}\Gamma^{IJKLM}\epsilon_1)H_{abh}F^{cdefg}{}_{h}X^{I}_c X^{J}_d X^{K}_e X^{L}_f X^{M}_g \nonumber\\[6pt]
    &-\tfrac{1}{2}(\epsilon_2^{T}\epsilon_1)F_{abcdef}(\Theta_c^{T}\Gamma_{IJ}\Theta_d)X^{I}_e X^{J}_f+\tfrac{1}{8}(\epsilon_2^{T}\Gamma^{IJKL}\epsilon_1)F_{abcdef}(\Theta_c^{T}\Gamma^{KL}\Theta_d)X^{I}_e X^{J}_f \nonumber\\
    &+\tfrac{1}{4}(\epsilon_2^{T}\Gamma^{I}\epsilon_1)F_{abcdef}(\Theta_c^{T}\Gamma^{IJK}\Theta_d)X^{J}_e X^{K}_f \nonumber\\[6pt]
    &+\tfrac{1}{6}i(\epsilon_2^{T}\Gamma^{IJKLM}\epsilon_1)H_{cdh}F_{abefgh}X^{I}_c X^{J}_d X^{K}_e X^{L}_f X^{M}_g \nonumber\\
    &+\tfrac{1}{6}i(\epsilon_2^{T}\Gamma^{KLMN}\epsilon_1)F_{abcegw}F_{dfhuvw} X^{I}_c X^{I}_d X^{J}_e X^{J}_f X^{K}_g X^{L}_h X^{M}_u X^{N}_v \nonumber\\
    &-\tfrac{1}{360}i(\epsilon_2^{T}\Gamma^{IJKLMNOP}\epsilon_1)F_{abcdew}F_{fghuvw} X^{I}_c X^{J}_d X^{K}_e X^{L}_f X^{M}_g X^{N}_h X^{O}_u X^{P}_v\,.
\end{align}
As one would expect, the first line can be written in terms of a gauge transformation
\begin{equation}
    [\delta_{\epsilon_2},\delta_{\epsilon_1}]\mathcal{A}_{ab}\supset D_t\Lambda_{ab}-2i(\epsilon_2^T\Gamma^I\epsilon_1)D_tH_{abc}X^I_c\,,
\end{equation}
where the gauge parameter is given by~\eqref{eq:gaugeparam_2}. The extra term appearing with the covariant derivative acting on $H_{abc}$ can be combined with the second line to give the equations of motion of $H$.
The term in the third line, namely
\begin{equation}
    -\tfrac{1}{60}i(\epsilon_2^{T}\Gamma^{IJKLM}\epsilon_1)H_{abh}F^{cdefg}{}_{h}X^{I}_c X^{J}_d X^{K}_e X^{L}_f X^{M}_g\eqqcolon H_{abc}\lambda_c\,,
\end{equation}
gives a gauge transformation $\delta_\lambda\mathcal{A}_{ab}$ as explained in the main text.
Lastly, the next two lines vanish upon symmetries of the gamma matrices, and the rest of the terms can be shown to vanish upon using the Filippov identities used throughout the text.

\section{Auxiliary-fermion multiplet completion}\label{app:auxiliary_fermion}

It may come as somewhat of a surprise that we were able to construct a manifestly supersymmetric theory in which the dynamical bosonic field $H_{abc}$ comes without a fermionic superpartner. We argued that this can be understood in terms of the fact that the situation is similar to 3D Chern-Simons gauge theory with $H_{abc}$ not propagating any dynamical degrees of freedom. For completeness, let us now analyze what the structure of the theory is when one introduces such a fermionic superpartner $\lambda_{abc}(t)$, which is also a 16-component real Majorana spinor. Guided by how one supersymmetrizes the Chern-Simons action, we extend the action of the theory by adding the following term\footnote{In fact, there exists a full family of extensions parametrized by $\kappa \in \mathbb{R}$ with action
\begin{equation*}
S_{\lambda,\kappa}=-\frac{i}{2}\!\int \! dt  \Bigl(\lambda_{abc}^T - \frac{\kappa}{2}F_{abcdef}\Theta_d^T \Gamma^{IJ}X^I_e X^J_f\Bigr) F_{abca'b'c'}\Bigl(\lambda_{a'b'c'} + \frac{\kappa}{2}F_{a'b'c'd'e'f'} \Gamma^{I'J'}\Theta_{d'}X^{I'}_{e'} X^{J'}_{f'}\Bigr)\,,
\end{equation*}
and $\delta_\epsilon \lambda_{abc} \to \delta_\epsilon \lambda_{abc}  -\frac{\kappa}{2} F_{abcdef}\delta_\epsilon (\Gamma^{IJ}\Theta_d X^I_e X^J_f), \delta_\epsilon H_{abc} \to \delta_\epsilon H_{abc} - \tfrac{i \kappa}{2} F_{abcdef}(\epsilon^T \Gamma^{JK} \Theta_d) X^J_e X^K_f$.
We will not discuss this family of extensions here, except for the case $\kappa=0$.}
\begin{equation}
S \supset S_\lambda=-\frac{i}{2} \!\int \! \mathrm{d}t\, \lambda_{abc}^T\, F_{abcdef}\, \lambda_{def}\,.
\end{equation}
The supersymmetry transformations for $H$ and the new field $\lambda$ read
\begin{equation}
\begin{aligned}
    \delta_\epsilon H_{abc} &= -i \epsilon^T \lambda_{abc} -\tfrac{i}{2} F_{abcdef}(\epsilon^T \Gamma^{JK} \Theta_d) X^J_e X^K_f\,,\\
    \delta_\epsilon \lambda_{abc}&=(D_t H_{abc}-\tfrac{i}{2}F_{abcdef}\Theta^T_d\Gamma^I\Theta_e X^I_f+\tfrac{1}{2}F_{abcdeh}H_{fgh}X^I_d X^J_e \, X^I_f X^J_g)\epsilon\,,
\end{aligned}
\end{equation}
with the transformations for $X,\Theta$ and $\mathcal A$ unaltered. One then realizes that the action is indeed invariant upon invoking the relation~\eqref{eq:squareid}. Since no Filippov identities are needed, this means that the extension can also be built for the $A_6$ model of Section~\ref{sec:A6}. The new term does not alter the existing equations of motion, while $\lambda$ has the trivial equation of motion
\begin{equation}
0=\lambda_{abc}\,.
\end{equation}
The closure of the commutator of two supersymmetry transformations acting on $X^I_a, \Theta_a, \mathcal{A}_{ab}$  generalizes straightforwardly, i.e. that one only has to adapt the equations of motion appearing  on the right hand sides. For the closure on the field $H_{abc}$ we obtain the known result~\eqref{eq:closure_H} without the need to impose the equations of motion of $H_{abc}$
\begin{equation}
\begin{aligned}
    [\delta_{\epsilon_2},\delta_{\epsilon_1}] H_{abc} \!&=
    2i(\epsilon_2^T \epsilon_1) D_t H_{abc} +  2i(\epsilon_2^T \Gamma^I \epsilon_1) F_{abcdef} X^I_d \big( X^J_e D_tX^J_f -\tfrac{i}{2} \Theta^T_e \Theta_f \big)\,.
\end{aligned}
\end{equation}
After a rather long and tedious calculation, the closure on the new field $\lambda_{abc}$ then gives
\begin{equation}
\begin{aligned}
    [\delta_{\epsilon_2},\delta_{\epsilon_1}] \lambda_{abc} \!&=
    \tfrac{i}{8}(\epsilon_2^T \epsilon_1) D_t \lambda_{abc}+\tfrac{i}{8}(\epsilon_2^T \Gamma^I\epsilon_1) \Gamma^I D_t \lambda_{abc}+\tfrac{i}{8}\tfrac{1}{4!}(\epsilon_2^T \Gamma^{IJKL}\epsilon_1) \Gamma^{IJKL} D_t \lambda_{abc}\,,
\end{aligned}
\end{equation}
where we have already omitted contributions from the equations of motion of $\Theta$ and $X$ that contribute at all three ranks.
Due to the trivial equation of motion of $\lambda$, also $D_t\lambda=0$ since $\lambda$ does not transform under the gauge transformation and hence the supersymmetry algebra closes on-shell. As one might expect, in order to derive this expression, one has to impose\footnote{For the special $A_6$ theory of Section~\ref{sec:A6} we do not expect this to be the case and closure should succeed without the need to impose this relation. Instead, $\lambda$ will transform under gauge transformations, such that $D_t \lambda$ is not automatically zero anymore. We leave the explicit calculation for future work.} the analogue of the Filippov identity (II) for $\lambda$
\begin{equation}
    \lambda_{[\uli{ab}}{}^{k} F_{k\uli{c}]mnpl}=0\,.
\end{equation}
Hence, this shows that we have constructed a bona fide  manifestly
supersymmetric theory containing  a supersymmetric pair $(H_{abc},\lambda_{abc})$ completing the multiplet. Consistently, integrating out the fermion $\lambda$, i.e. setting $\lambda=0$ in the action, we recover the theory discussed before.

\newpage
\bibliography{references}
\bibliographystyle{utphys}

\end{document}